\documentstyle[12pt]{article}
\begin{document}

%\begin{center}
{\noindent \Large \bf  On possibility to observe new fundamental
forces in open quantum systems}\\
%\end{center}

\begin{center}
\bf B.F. Kostenko\footnote{Joint Institute for Nuclear Research,
 Dubna; e-mail:bkostenko@jinr.ru}, \bf M.Z. Yuriev\footnote{ EuroFinanceGroup, Moscow}
\end{center}

\bigskip

%\begin{center}
%\bf  Abstract
%\end{center}

{\it A possibility and peculiarities of registration of new
fundamental forces in open quantum systems are discussed. As a
possible example, variations of decay rates of radioactive
elements reported in scientific literature are considered in
detail.}

\bigskip
\bigskip
\bigskip

%\begin{center}
{\noindent  \bf  1. INTRODUCTION}
%\end{center}
\smallskip

During last three decades the theory of open quantum systems was
developed  in several directions. Among  new approaches arising in
this  period are the Lindblad equation for density matrix,
operator generalization of the Ito integral, different methods for
descriptions of quantum systems interacting with the measuring
device, and others. A range of physical applications of the
mathematical formalism was also extended appreciably. It includes
new attempts to describe the quantum measurements, theory of
tunneling processes under dissipation conditions, solution of
different problems concerning models of quantum computers and
quantum lines of communications.

It is necessary, however, to note a fact (practically escaped
researchers' attention) that open quantum systems can also serve
as peculiar detectors of new fundamental forces that can be
feasible at present only by means of the systems of this genus. As
an example, we refer to the quantum gravitational interference
destruction mechanism  leading to an equation of the Lindblad
type$^{(1)}$. In the general case, the forces in question can be
any rather weak ones which act on the quantum system from the part
of a macro- or megasystem. Corresponding tests  have features
discriminating them from the usual accelerator experiments and,
therefore, there is a risk to interpret the appearance of new
interactions as drawbacks of an experiment itself and,
consequently, to reject it.

This paper is devoted to consideration of some general properties
of open quantum systems in context of the possibility to search
the new fundamental interactions, as well as to discussion of some
experiments in which such interactions could be already seen.

\bigskip
%\newpage
%\begin{center}
{\noindent \small \bf 2. JOINT CONSIDERATION OF QUANTUM SYSTEM AND
MEASURING INSTRUMENT}
%\end{center}
\smallskip

As is well known$^{(2)}$, the Schr\"odinger and von Neumann
equations for closed quantum systems are intimately connected with
continuous projective representations of the Galilean
group\footnote{ Generalization of this device onto the Lorentz
group is well-known too$^{(3)}$, but here we restrict ourselves to
consideration of a more simple nonrelativistic case. }. The action
of the one-parametric sub-group of temporal shifts on the set of
states of a quantum system is described by the formula
\begin{equation}\label{1}
\varrho_t = e^{-iHt}\varrho_0 e^{iHt},
\end{equation}
where $H$ is Hamiltonian of the system. From this, after the
formal differentiation over $t$, we obtain the von Neumann
equation which gives rise to the Schr\"odinger one for the pure
states, $\varrho_t = |\psi_t \rangle \langle\psi_t|$. Interactions
of the quantum system with its environment results in breakdown of
the symmetry with regard to some transformation of the  Galilean
group, and this circumstance turns the theory of representations
of groups into an important instrument for the open quantum system
investigations too. The mathematical apparatus elaborated by now
(in particular, the method of dynamic semigroup (see Refs. 4 and
5)) takes into account, in an explicit form, only the breakdown of
symmetry for the sub-group of the temporal shifts. In what
follows, we shall formulate a model for which  the spatial
rotation symmetry breakdown will play an essential role.

Any interactions of quantum system,  S, with its environment come
to light due to their influence on the probability distribution of
measurement results. Let us denote by $\mu_{\rho} (\chi) d\chi$ a
probability of some observable registration  by means of a
measuring instrument, M, in the interval $(\chi, \chi+d\chi)$, if
the quantum system is in a state $\rho =\sum_i p_i |\psi_i \rangle
\langle \psi_i|$. In the general case the probability distribution
$\mu_{\rho} (\chi) d\chi$ may be singular and the variable $\chi$
several-dimensional, as is the case with correlative spin
experiments of Einstein-Podolsky-Rosen type. According to the
existing theory, $$\mu_{\rho} (\chi) d\chi = Tr [\rho M(d\chi)],
$$ where $M(d\chi)$  is the positive operator valued measure
describing the action of M on S. In the Dirac notations $M(d\chi)=
|\chi \rangle \langle \chi| d\chi,$ where the orthogonality
condition, $$\langle\chi|\chi^{\prime}\rangle = \delta(\chi
-\chi^{\prime}),$$ is considered neither necessary nor realistic
one, but the resolution of identity, $$ \int |\chi \rangle \langle
\chi| d\chi =1,$$ is supposed to take place$^{(6)}$.

As a result of some Galilean group transformation, $g$, the
density matrix and the measuring instrument are changed:
\begin{equation}\label{2}
\rho_g = U_g \rho U_g^{\dagger}, \qquad M_g(d\chi) = V_g M(d\chi)
V_g^{\dagger}.
\end{equation}
Here, in accordance with the Wigner theorem, $U_g$ and $V_g$ are
unitary or antiunitary operators acting in the Hilbert space of
states of system S. Equations (\ref{2}) generalize formula
(\ref{1}) for the case of a simultaneous consideration of S and M
systems. Therefore, for the sub-group of  temporal shifts, they
replace  the Schr\"odinger equation. Probability distribution
$\mu_{\rho} (\chi) d\chi$ will be invariant under $g$
transformation if and only if $$ U_g = V_g .$$

Actually, measuring instrument M is a compound system consisting
of closed  and open subsystems, M $=$ M$_c \;+$ M$_o$.  In the
beginning of the measurement process the closed part M$_c$
interacts  only with S and, after receiving a sufficient energy,
it starts interacting with M$_o$. That makes sense to speak about
the unitary transformation, $V_g$, only for $M_c$, which is the
most sensitive part of M.

Breakdown of the invariance with respect to the Galilean group can
be caused by interactions both S and M$_c$ with environment. In
this case, operators $U_g $ and $V_g$ may be nonunitary, or, in
the general way, equations (\ref{2}) cannot be applicable at all,
since the Wigner theorem conditions are violated. Theoretically,
new fundamental forces may reveal themselves not only by means of
the true  Galilean evolution of M$_c$ and a distorted one of S
(the traditional conception of the measurement), but also in the
inverted form, when M$_c$ is distorted and S is not, or due to S
and M$_c$ simultaneous distortion. Observable breakdown of the
invariance properties of M$_o$  with respect to the Galilean group
are usually accepted as incorrect performance of the experiment.

Joint consideration of the quantum system  and the measuring
instrument, though it seems to be  trivial, in fact is more
general than a traditional approach based on a separate
description of the quantum system alone. In particular, it permits
to generalize in a reasonable way the very concept of interaction.
For example, quantum cryptography is based on the fact that
$\mu_{\rho} (\chi)d\chi $ is altered because of the illegal access
results in replacement of  $\rho_t =  U_t \rho U_t^{\dagger}$  by
some other transformation, although a change of energy of the
quantum system not necessarily takes place after the manipulations
of a disturber are completed (7--10). It is reasonable to
designate these interactions by a special name, e.g. {\it
informational}, to distinguish them from the standard potential
ones. Such interactions manifest themselves as an {\it extra
factor} (which cannot be calculated using the Schr\"odinger
equation) acting on the probability distribution $\mu_{\rho}
(\chi) d\chi$.

Special examples of the informational interactions are the noted
quantum Zeno and anti-Zeno effects (see Refs. 11 and 12). In these
cases controllable changes of a measuring instrument which
``observes'' the state of an unstable system can lead to changing
the decay probability, $\mu_{\rho} (t)dt$, of the system.

\bigskip

%\begin{center}
{\noindent \small \bf  3. CONCRETE EXPERIMENTAL EXAMPLES}
%\end{center}
\smallskip

Now let us consider some papers in which variations of decay rates
of radioactive elements have been reported and, presumably,
unknown interactions have been registered. These unusual
experiments can serve only as a possible illustration to our
previous reasoning since the announced results are a challenging
diversion from the standard lore in the subject and still demand
further careful examination.

In paper$^{(13)}$, in spite of the efforts undertaken for
elimination of temperature, man-caused and other effects on the
recording equipment, changes of $\beta$-decay rate, with periods
24 hours and 27 days, were observed at two laboratories 140 km
apart (INR RAS, Troitsk, $^{60}$Co, and JINR, Dubna, $^{137}$Cs).
Extremum deviations of count rate ($ 0.7\%$ for $^{60}$Co and
$0.2\%$ for $^{137}$Cs) from the statistical average took place
for the both laboratories when they were oriented properly along
the three definite directions established in the outer space.
Bursts of count rate of beta-radioactive sources during long-term
measurements, similar to data of Ref. 13, were also reported in an
independent paper$^{(14)}$.

Series of papers devoted to dependence of $\alpha$-activity, as
well as different macroscopic fluctuations, on the cosmological
factors was published by Prof. S.E.~Shnol et al. in Russian
scientific journals (see Ref. 15 for details and the
bibliography). Here we give only a brief review of the data from
Refs. 16 and 17. In these studies a phenomenon of a deviation of
probability distributions from the expected Poisson one was
established. The measurements were carried out in fixed with
respect to the Earth's surface laboratories during 5 minute time
intervals. Non-randomness of repetitions  of the shape of the
observed distributions was also established at the regular time
intervals. In short, the main results are the following:
\begin{enumerate}
\item Re-appearance of the same form of a probability distribution
took place most likely in the nearest  interval of observation.
\item There was a reliable growth of probability of the same form to re-appear
after 24 hours, 27 days, and one year.
\item Synchronous measurements of the form carried out in
different laboratories showed that for distances less than 100 km
about 60$\%$ pairs of the distributions had the same form.
Probability to observe similar distributions turned out to be high
also for measurements on a research ship in the Indian Ocean and
in a remote laboratory near Moscow, which were in the same time
zone.
\end{enumerate}

These data, in the case of their conformation, will almost
undoubtedly testify against the invariance of the radioactive atom
(and/or detector) properties with respect to spatial rotations.
Seemingly, just this conclusion follows at once from the both
principal papers Refs. 13 and 16. A discrepancy demanding further
consideration is only elucidation of a possible source of the
influence on  S $+$ M$_c$. Indeed, according to the experiments of
Shnoll at al., it is natural to connect the observed effect with
the influence of the nearest cosmic environment, such as the Sun
and the Moon. Authors of Ref. 13 explain a dependence of
$\beta$-decay rate by the mutual orientation of S system and
unknown cosmic field directed toward the Constellation Hercules.

\bigskip
%\newpage
%\begin{center}
{\noindent \small \bf  4. NEW QUANTUM NUMBERS}
%\end{center}
\smallskip

From the preceding considerations it is possible to give the
following explanation of the observed phenomena. Generators of the
spinor representation of the rotation sub-group,
\[ \sigma ^z  = \left( {\begin{array}{*{20}c}
   1 & 0  \\
   0 & { - 1}  \\
\end{array}} \right), \qquad \sigma ^ +  =  \sigma ^ x + i  \sigma ^ y   =
 \left( {\begin{array}{*{20}c}
   0 & 1  \\
   0 & 0  \\
\end{array}} \right), \qquad  \sigma ^ -  =  \sigma ^ x - i  \sigma ^ y  =
\left( {\begin{array}{*{20}c}
   0 & 0  \\
   1 & 0  \\
\end{array}} \right),
\]
can be factorized by means of the relations:
\[
\sigma ^ +   = a^\dag  b, \qquad  \sigma ^ -   = b^\dag  a,
\]
where
\[
a=|0\rangle \langle +|, \qquad b=|0\rangle \langle -|,
\]
and $|0\rangle $  is the vacuum state. Sign $^\dag$ denotes the
hermitian conjugation. Actually, we introduce in such a way the
birth and annihilation operators for atoms {\it ready}  and {\it
unready} to decay (atoms of the type $a$ and $b$,
correspondingly). They satisfy the fermion anticommutative
relations,
\[
aa^\dag + a^\dag a = 1,  \qquad bb^\dag + b^\dag b = 1 ,
\]
which can be also interpreted as resolutions of identity in the
Fock spaces for particles of types  $a$ and $b$.

From the physical point of view, the undertaken factorization
implies the definition of  new quantum numbers, $$n_a =  a^\dag a,
\qquad n_b =  b^\dag b,$$ which correspond to probabilities of
atom to decay and to survive, correspondingly. Since $ \sigma ^z
= n_a - n_b$ and eigenvectors of operators $n_a$ and $n_b$
coincide with eigenvectors of $ \sigma ^z $,  we shall call the
property of the radioactive atom to decay by {\it quasi-spin}.

Under spatial rotations, the ability of the system of the atom
plus the measuring instrument to demonstrate the decay, in the
general case, are changed: $$ Tr[\rho n_a] \rightarrow Tr[U \rho
U^\dagger \; V n_a V^\dagger].$$ If, e.g., we take an atom
completely ready to decay, $ \rho_+ = |+\rangle \langle +|, $ and
rotate it relatively to the fixed instrument, than corresponding
transformations appear as follows: $$V=1; \qquad U\;|+\rangle =
\alpha |+\rangle + \beta |-\rangle, $$ where $|\alpha|^2 +
|\beta|^2 =1.$ Thus the spatial rotations lead, in our model, to
changing the probability to observe the decay by the factor
$|\alpha|^2.$ We suggest that there are fixed directions in the
cosmic space such that atoms are the most unstable if their
quasi-spins are oriented along them.

Positive hints at the existence of the quasi-spin polarization
follow from the experiment. Indeed, although the experiment did
not reveal noticeable correlations between changes of
$\beta$-decay rate for two different radioactive sources,
$^{60}$Co and $^{137}$Cs, the sources demonstrated a systematic
increase of nuclear decay rate at their quite certain, different
for different sources, spatial orientations$^{(18)}$. But if even
a state of the polyatomic system is not polarized, the existence
of the ``readiness-to-decay'' quantum number can be observed
because of its influence, especially significant for completely
unpolarized atomic ensembles, on the quantum statistics. Besides,
models which will be considered in the next section show that the
energy of an unstable atom should depend on its orientation
regarding to a pseudo-vector of current interacting with the atom
by means of a new, ``the fifth'' force.

\bigskip
%\begin{center}
{\noindent \small \bf 5. MODELS OF INTERACTIONS}
%\end{center}
\smallskip

The ``readiness to decay'' of atoms can be changed by an external
field, $\varphi$, carrying this quantum number. Our
non-relativistic consideration does not forbid us to introduce the
following interaction in the spirit of the Lee model$^{(19)}$: $$
H_{int} = \frac{\lambda}{\sqrt{N}} \sum_{i=1}^N (\varphi
a^\dagger_i b_i + b^\dagger_i a_i \varphi^\dagger), $$ where $N$
is a number of radioactive atoms and  $\lambda$ is a coupling
constant. Evidently, this interaction preserves the total number
of atoms and  the total ``readiness to decay''. In this model,
experimentally observed variations of nuclear decay rates could be
a consequence of exchange between radioactive atoms on the Earth
and the Sun by quanta of the $\varphi$ field (the corresponding
Feynman graph is quite obvious).

Interactions between radioactive atoms can be  also written in the
form of the Fermi 4-particle interaction, i.e. as
``current$\times$current''\footnote{ Actually, this means that we
consider an effective field theory which can be obtained after
fixing a proper energy cut-off and taking the path integral over
the high-frequency fields (see, e.g., Ref. 20).}. As far as the
current components here are the Pauli matrices, $\sigma_x, \;
\sigma_y, \;\sigma_z$, the interaction invariant with regard to
the spatial rotations can be written in the form: $$ v = v(r) \;
(\vec \sigma_1 \cdot \vec \sigma_2). $$ It does not modify the
total quasi-spin,  $\vec S =\vec \sigma_1 + \vec \sigma_2 $, of
two interacting atoms. The obtained potential resembles the spin
dependent nucleon potential, and the theory of nuclear forces
prompts one more possible potential,
 $$  u = u(r) \;
[3(\vec \sigma_1 \cdot \vec n) (\vec \sigma_2 \cdot \vec n) -
(\vec \sigma_1 \cdot \vec \sigma_2) ],$$ which preserves  $\vec
S^2 .$  Apparently, assumptions of this model could be tested and,
if needed,  $v(r),$ $u(r)$, could be established  in space ship
experiments.

It is clear, a current acting upon the radioactive atom may be not
only the quasi-spin of other atoms, but a pseudovector of a
different nature too. In this connection, an assumption$^{(21)}$
that the pseudovector of current of the so-called light
monopole$^{(22)}$ can be an effective catalyst of the weak decays
is of interest. Another possible source is enigmatic dark matter
and dark energy which prevail in the universe according to the
modern astrophysical data. It is also possible to identify the
field $\varphi$ with Goldstone excitations corresponding to
breaking the rotational symmetry of the space.

%\smallskip
\bigskip
%\newpage
%\begin{center}
{\noindent \small \bf 6. ``PREGEOMETRY''}
%\end{center}
\smallskip

As it was pointed out above, the operators describing the
``readiness to decay'' can be found with the help of  the
factorization of spatial rotation sub-group generators. In their
turn, these generators appear naturally if we use the Dirac
(non-commutative) factorization of the spatial metrics: $$ds^2 =
\eta_{\alpha \beta} dx^\alpha dx^\beta = \frac{1}{2}\sigma_\alpha
\sigma_\beta dx^\alpha dx^\beta,$$ where we take into account the
anti-commutative properties of the Paili matrices: $$\sigma_\alpha
\sigma_\beta  + \sigma_\beta \sigma_\alpha = 2 \delta_{\alpha
\beta}.$$ This prompts us to an idea  that the described mechanism
of the decay control can be related to general properties of the
space itself and, therefore, be universal. Such a strong
assumption is also supported by a resemblance  of the nuclear
decay rate probability distributions to fluctuations of
biochemical reaction rates, which are observed concurrently at the
same spatial location$^{(16)}$. It is also close enough to an idea
of Misner, Thorn and Wheeler, who speculated about existence of a
connection between geometry and classical two-valued
logic$^{(22)}$. The connection between geometry and logics is
caused in our model by dual properties of the Pauli matrices
$\sigma_x, \sigma_y, \sigma_z$. Indeed, they are transformed as
components of the axial vector, on the one hand, and they operate
in the space of the quasi-spin components, on the other hand. But,
according to our interpretation, the quasi-spin controls the truth
values of statements about realization of a fixed result of two
possible outcomes. Besides, the measurement process plays an
important role. It turns a closed quantum system into an open one
(in the sense  of the informational interactions defined above).

Of course, the exciting theoretical possibilities considered in
the present paper will be urgent only in the case of the final
confirmation of the data presented in Refs. (13--17). Therefore,
the most topical task now is checking the results of the
above-mentioned papers and carrying-out analogous experiments
under altered conditions like setting the device with radioactive
sources on a rotating platform, in underground or cosmic
laboratories, etc. To stimulate interest to this poorly understood
problem was one of the aims of our paper.

\newpage

%\begin{thebibliography}{21}
%\begin{sloppypar}
\bigskip
\bigskip
\bigskip
{\noindent \small \bf REFERENCES} \smallskip
\begin{enumerate}
\item  J. Ellis, S. Mohanty, and D.V. Nanopoulos, {\it
Phys. Lett. B} {\bf 221}, 113 (1989); {\bf 235}, 305 (1990).

\item E. In\"onu and E.P. Wigner, {\it Nuovo Cim.} {\bf
9}, 705 (1952).

\item   E.P. Wigner, {\it Ann. Math.} {\bf 40}, 149 (1939);
{\it Nucl. Phys. B (Proc. Suppl.)} {\bf 6}, 9 (1989).

\item   D.~Zwanziger, {\it Phys. Rev.} {\bf 131}, 2818
(1963).

\item   L.~ Lanz, L.A.~Lugiato, and G.~Ramella, {\it Int. J.
Theor. Phys.} {\bf 8},  341 (1973).

\item  C.W. Helstrom, {\it Quantum detection and estimation
theory} (Academic Press, N.Y., 1976).

\item  C.H.~Bennett, G.~Brassard, and A.K.~Andekert, {\it
Scientific Amer.}, {\bf 267},  50 (1992).

\item  C.H.~Bennett, {\it Phys. Rev. Lett.} {\bf 68}, 3121
(1992).

\item  A.K.~Ekert, J.~Parity, P.~Tapster, and G.~Palma,
{\it Phys. Rev. Lett.} {\bf 69}, 1293 (1992).

\item  H.-K.~Lo and H.F.~Chau, {\it Science}, {\bf 283},
2050 (1999).

\item  B.~Misra and E.C.G.~Sudarshan, {\it J. Math. Phys.}
{\bf 18},  756 (1977).

\item  A.G.~Kofman and G.~Kurizki, {\it Phys. Rev. A} {\bf
54},  R3750 (1996).

\item  Yu. A. Baurov, A.A. Konradov, V.F. Kushniruk, E.A.
Kuznetsov, Yu.G. Sobolev, Yu. V. Ryabov, A.P. Senkevich,  and S.V.
Zadorozsny,  {\it Mod. Phys. Lett. A} {\bf 16},  2089 (2001).

\item  A.G. Parkhomov, {\it Int. J. of Pure and Appl.
Phys.}  {\bf 1},  119 (2005).

\item  S.E. Shnoll et al. e-prints: physics/0603029,
0602017, 0504092, 0501004, 0412152, 412007.

\item   S.E. Shnoll,  V.A. Kolombet,  E.V.  Pozharski,
T.A. Zenchenko,  I.M. Zvereva, and  A.A. Konradov, {\it
Physics-Uspehi} {\bf 162}, 1129 (1998).

\item   S.E. Shnoll,  T.A. Zenchenko,  K.I. Zenchenko,
 E.V. Pozharski,  V.A. Kolombet, and A.A. Konradov,  {\it Physics-Uspehi}, {\bf
 43},  205 (2000).

\item  Yu. A. Baurov, private communication.

\item  T.D.~Lee, {\it Phys. Rev.} {\bf 95},  1329 (1954).

\item  J. Polchinski, ``Effective Field Theory and the
Fermi Surface'', Lectures presented at TASI 1992, e-print:
hep-th/9210046.

\item  A.A.~Rukhadze, L.I.~Urutskoev, and D.V.~Filippov,
{\it Bull. Lebedev Phys. Inst.} {\bf 1} (2004).

\item  G.~Lochak, {\it Int. J. Theor. Phys.} {\bf 24},
1019 (1985).

\item  Ch.W. Misner, K.S. Thorn, and J.A. Wheeler, {\it
Gravitation} (Freeman and Company, San Francisco, 1973).

\end{enumerate}

%\end{thebibliography}

\end{document}